\documentclass[aps,preprint,showpacs,showkeys]{revtex4}
\usepackage{amsfonts}
\usepackage{amsmath}
\usepackage{amssymb}
\usepackage{graphicx}

\begin{document}

\title{On the Dirac equation with PT-symmetric potentials in the presence of position-dependent mass}\author{L.B. Castro}\email[ ]{benito@feg.unesp.br}
\affiliation{Departamento de F\'{\i}sica e Qu\'{\i}mica,
Universidade Estadual Paulista, 12516-410 Guaratinguet\'{a}, S\~ao
Paulo, Brazil} 

\begin{abstract}

The relativistic problem of fermions subject to a PT-symmetric potential in the presence of position-dependent mass is reinvestigated. The influence of the PT-symmetric potential in the continuity equation and in the orthonormalization condition are analyzed. In addition, a mis\-con\-cep\-tion diffused in the literature on the interaction of neutral fermions is clarified.

\end{abstract}

\maketitle

After the pioneer work by Bender and Boettcher \cite{ben}, the so-called PT-symmetric systems have been studied widely because of their intrinsic interest and also for applications in different research field, as for example, in the study of nuclear physics \cite{n1}-\cite{n2}, quantum field theories \cite{tq1}-\cite{tq3} and electromagnetic wave traveling in a planar slab waveguide \cite{rus}. In recent years, some authors have extended the research fields for the PT-symmetric quantum systems with a constant mass to the relativistic PT-symmetric position-dependent effective mass quantum systems. It is usually expected that, in the relativistic case, the ordering ambiguity of the mass and operators, which is present in the non-relativistic case, should disappear. Systems with position-dependent mass have been found to be very useful in the description of electronic properties of, for instance, compositionally graded crystals \cite{gell}, quantum dots \cite{serra}, quantum liquids \cite{arias}, etc. Many problems of fermions interacting with PT-symmetric potentials in presence of position-dependent mass have been reported in the literature \cite{sin}-\cite{chun}. Shina and Roy \cite{sin} investigated the one-dimensional solvable Dirac equation with non-Hermitian scalar and pseudoscalar interactions (wherein the scalar interaction plays the role of a position-dependent mass), possessing real energy spectra. In Refs. \cite{chun3}-\cite{mus2}, the authors proposed a scheme to construct a PT-symmetric potential with a real relativistic energy spectrum in the setting of the position-dependent effective mass Dirac equation in 1+1 dimensions. In the case of the position-dependent mass distribution with linear and inversely linear form, the bound state solutions of the effective mass Dirac equation for a singular PT-symmetric potential have been studied \cite{chun3}. By using the method proposed in Ref. \cite{chun3}, the relativistic problem of neutral fermions subject to PT-symmetric trigonometric potential in 1+1 dimensions has been investigated \cite{chun4}. Following up of the work \cite{chun3}, a new method has been proposed to construct the exactly solvable PT-symmetric potentials within the framework of the (1+1)-dimensional position-dependent effective mass Dirac equation with the vector potential coupling scheme \cite{chun2}. Mustafa and Mazharimousavi \cite{mus} investigated the spectrum of the (1+1)-Dirac equation with position-dependent mass and complexified PT-symmetric Lorentz scalar interactions. In Ref. \cite{chun} the bound state solutions of the effective mass Dirac equation in 1+1 dimension in presence of a position-dependent double-well-like mass distribution and a symmetric mass distribution is studied. In all above research, do not analyze the four-current or the orthonormal condition.

In this present paper the relativistic problem of fermions subject to a PT-symmetric potential in the presence of position-dependent mass is reinvestigated. The four-current in the presence of interaction for fermions with position-dependent mass is analyzed and the condition for a conserved four-current is obtained. An orthogonality criterion for the eigenspinors is founded. This kind of analysis has already a precedent in the Duffin-Kemmer-Petiau (DKP) equation \cite{yoo1}-\cite{yoo4}. These results are used to show not only that the presence of PT-symmetric potential does not lead to a conserved current, but also to show that the orthogonality criterion is consistent depending on the suitable boundary condition. In addition, it is also shown that a pseudoscalar potential in (1+1) dimension is associated with interaction of neutral fermions, as opposed to what was done in Refs. \cite{chun4} and \cite{chun}.

The Dirac equation for a free fermion with position-dependent mass $M(\vec{r})$ is given by (with units
in which $\hbar =c=1$)%
\begin{equation}
\left( i\gamma ^{\mu }\partial _{\mu }-M(\vec{r})\right) \psi =0  \label{dkp}
\end{equation}%
\noindent where the matrices $\gamma^{\mu }$\ satisfy the algebra%
\begin{equation}
\gamma ^{\mu }\gamma^{\nu }+\gamma ^{\nu }\gamma^{\mu }=2\eta^{\mu\nu} \label{beta}
\end{equation}%
\noindent and the metric tensor is $\eta^{\mu \nu
}=\,$diag$\,(1,-1,-1,-1)$. The algebra expressed by (\ref{beta})
generates a set of 16 independent matrices.

A well-known conserved four-current is given by
\begin{equation}
J^{\mu }=\bar{\psi}\gamma ^{\mu }\psi   \label{corrente}
\end{equation}%
\noindent where the adjoint spinor $\bar{\psi}$ is given by%
\begin{equation}
\bar{\psi}=\psi ^{\dagger }\gamma ^{0}  \label{adj}
\end{equation}%
with%
\begin{equation}
\left(\gamma ^{0}\right)^{\dagger}=\gamma^{0}  \label{etamu}
\end{equation}%
in such a way that $\left( \gamma^{0}\gamma ^{\mu }\right) ^{\dagger
}=\gamma ^{0}\gamma ^{\mu }$ (the matrices $\gamma ^{\mu }$ are
Hermitian with respect to $\gamma ^{0}$). The time component of this
current is positive definite but it may be interpreted as a
charge density. Then the normalization condition can be expressed as
\begin{equation}
\int d\tau \,\bar{\psi}\gamma ^{0}\psi = 1  \label{norm}
\end{equation}%

With the introduction of interactions, the Dirac equation (\ref{dkp}) can be written as%
\begin{equation}
\left( i\gamma ^{\mu }\partial _{\mu }-M(\vec{r})-\mathcal{V}(\vec{r})\right) \psi =0
\label{dkp2}
\end{equation}%
where the more general potential matrix $\mathcal{V}(\vec{r})$ is written in terms of
16 linearly independent matrices. In the
presence of interactions $J^{\mu }$ satisfies the equation%
\begin{equation}
\partial _{\mu }J^{\mu }+i\bar{\psi}\left( \mathcal{V}-\gamma ^{0}\mathcal{V}^{\dagger }\gamma
^{0}\right) \psi =0 \label{corrent2}
\end{equation}%
Thus, if $\mathcal{V}$ is Hermitian with respect to $\gamma ^{0}$, i.e.
\begin{equation}\label{ps}
    (\gamma^{0}\mathcal{V})^{\dagger}=\gamma^{0}\mathcal{V}
\end{equation}
\noindent then four-current will be conserved. A similar procedure was already reported for DKP equation in \cite{yoo1}-\cite{yoo4}. The potential matrix $\mathcal{V}$ can be
written in terms of well-defined Lorentz structures. For (3+1) dimensions there are a scalar, a pseudoscalar, a vector, a pseudovector and a tensor terms \cite{grei}. In this stage, it is worthwhile to mention that the Dirac equation with position-dependent mass is equivalent to a Dirac equation for massless fermions with a scalar potential.

If the terms in the potential $\mathcal{V}$ are time-independent one can write $\psi (%
\vec{r},t)=\phi (\vec{r})\exp (-iEt)$, where $E$ is the energy of
the fermion,
in such a way that the time-independent Dirac equation becomes%
\begin{equation}
\left[ \gamma ^{0} E +i\gamma^{i}\partial _{i} - M-\mathcal{V} \right] \phi =0  \label{DKP10}
\end{equation}%
In this case \ $J^{\mu }=\bar{\phi}\gamma ^{\mu }\phi $ does not
depend on time, so that the spinor $\phi $ describes a stationary
state. Eq. (\ref{DKP10}) for the characteristic pair $(E_{k},\phi _{k})$
can be
written as%
\begin{equation}
 E_{k} \gamma ^{0}\phi_{k}+ i \overrightarrow{\partial _{i}}\gamma ^{i}\phi _{k}-M\phi _{k} -\mathcal{V}\phi _{k}=0  \label{orto1}
\end{equation}%
and its adjoint form, by changing $k$ by $k^{\prime }$, as%
\begin{equation}
E_{k^{\prime }}
\bar{\phi}_{k^{\prime }}-i\bar{\phi}_{k^{\prime }}\gamma
^{i}\gamma ^{0}\overleftarrow{\partial _{i}} -M\bar{\phi}%
_{k^{\prime }}\gamma ^{0}-\bar{\phi}_{k^{\prime }}\gamma ^{0}\mathcal{V}^{\dagger }=0
\label{orto2}
\end{equation}%
By multiplying (\ref{orto1}) from the left by $\bar{\phi}_{k^{\prime
}}$ and
(\ref{orto2}) from the right by $\gamma ^{0}\phi _{k}$ leads to%
\begin{equation}
E_{k} \bar{\phi}_{k^{\prime
}}\gamma^{0}\phi _{k}+i\bar{\phi}_{k^{\prime }}\gamma ^{i} \overrightarrow{%
\partial _{i}}\phi _{k}-M\bar{\phi}%
_{k^{\prime }}\phi _{k}-\bar{\phi}_{k^{\prime }}\mathcal{V}\phi _{k}=0  \label{orto3}
\end{equation}%
and%
\begin{equation}
E_{k^{\prime }}\bar{\phi}_{k^{\prime }}\gamma ^{0}\phi _{k}-i\bar{\phi}_{k^{\prime
}}\gamma ^{i} \overleftarrow{\partial _{i}} \phi _{k}-M%
\bar{\phi}_{k^{\prime }}\phi _{k}-\bar{\phi}%
_{k^{\prime }}\gamma ^{0}\mathcal{V}^{\dagger}\gamma ^{0}\phi
_{k}=0 \label{orto4}
\end{equation}%
respectively. Subtracting (\ref{orto4}) from (\ref{orto3}) and
considering
that the spinors fit boundary conditions such that%
\begin{equation}
\int d\tau \,\partial _{i}\left( \bar{\phi}_{k^{\prime}}\gamma ^{i}\phi
_{k}\right) =0  \label{orto5}
\end{equation}%
\noindent and that $(\gamma^{0}\mathcal{V})^{\dagger}=\gamma^{0}\mathcal{V}$, one gets%
\begin{equation}
\left( E_{k}-E_{k^{\prime }}\right) \int d\tau \,\bar{\phi}_{k^{\prime}}\gamma
^{0}\phi _{k}=0  \label{orto6}
\end{equation}%
Eq. (\ref{orto6}) is an orthogonality statement applying to the Dirac
equation. Any two stationary states with distinct energies and
subject to suitable boundary conditions are orthogonal in the sense
that
\begin{equation}
\int d\tau \,\bar{\phi}_{k^{\prime}}\gamma ^{0}\phi _{k }=0,\quad \mathrm{for}%
\quad E_{k}\neq E_{k^{\prime }}  \label{orto7}
\end{equation}%
In addition, in view of (\ref{norm}) the spinors $\phi _{k}$ and
$\phi_{k^{\prime }}$ are said to be orthonormal if%
\begin{equation}
\int d\tau \,\bar{\phi}_{k^{\prime}}\gamma ^{0}\phi _{k}= \delta
_{E_{k^{\prime}}E_{k}}  \label{orto8}
\end{equation}

\noindent Hence, the orthonormal condition is guaranteed if and only if (\ref{ps}) and (\ref{orto5}) are satisfied.

Let now consider the (1+1)-dimensional time-independent Dirac equation with position-dependent mass. In this case the potential matrix $\mathcal{V}$ can be written as
 \begin{equation}\label{gp}
    \mathcal{V}=\gamma^{0} V_{t}+\gamma^{1}V_{sp}+V_{s}-i\gamma^{5}V_{p}
 \end{equation}

 \noindent This is the most general combination of Lorentz structures because there are only four linearly independent $2\times2$ matrices. The subscripts for the terms of the potential denote their properties under a Lorentz transformation: $t$ and $sp$\, for the time and space components of the 2-vector potential, $s$ and $p$ for the scalar and pseudoscalar terms, respectively. Considering only the time component of the 2-vector potential, $\mathcal{V}$ is the form
 \begin{equation}
 \mathcal{V}=\gamma ^{0}A(x)\label{pot}
 \end{equation}%

 \noindent such a way that the time-independent Dirac equation becomes
 \begin{equation}\label{de}
    \left[\gamma^{0}E+i\gamma^{1}\partial_{x}-M(x)-\gamma^{0}A(x)\right]\phi=0
 \end{equation}

 \noindent To have an explicit expression for the $\gamma^{0}$ and $\gamma^{1}$ matrices one can choose $2\times2$ Pauli matrices which satisfy the same algebra. One can explicitly choose $\gamma^{0}=\sigma_{1}$ and $\gamma^{1}=-i\sigma_{2}$. The spinor can be written as $\phi^{T}=(\phi_{+},\phi_{-})$ in such a way that the Dirac equation for a fermion constrained to move along the $X$-axis gives rise to two coupled first-order equations
 \begin{equation}\label{de1}
    i\frac{d\phi_{+}}{dx}+\left[ E- A(x) \right]\phi_{+}-M(x)\phi_{-}=0
 \end{equation}
 \begin{equation}\label{de2}
    -i\frac{d\phi_{-}}{dx}+\left[ E- A(x) \right]\phi_{-}-M(x)\phi_{+}=0
 \end{equation}

 \noindent This two coupled differential equations are the same as obtained in Refs. \cite{chun3}-\cite{chun2}. The vector potential is imposed in the form of a non-Hermitian imaginary potential
 \begin{equation}\label{pts}
    A(x)=\frac{i}{2}\, \frac{1}{M(x)}\,\frac{dM(x)}{dx}.
 \end{equation}

 \noindent A potential $A(x)$ is said to possess PT symmetry if the relation $A(-x)=A^{*}(x)$ or $A(\xi-x)=A^{*}(x)$ exists under the transformation of $x\rightarrow-x$ (or $x\rightarrow\xi-x$) and $i\rightarrow-i$, where $\mathcal{P}$ denotes parity operator (space reflection) and $\mathcal{T}$ denotes time reversal. The potential (\ref{pts}) shows $A(-x)=A^{*}(x)$ because $M(x)$ is invariant under the parity operation ($M(-x)=M(x)$), thus, this complex potential possesses PT symmetry.

In this stage, it is useful to analysis the current for this kind of quantum system. Consider the condition (\ref{ps}) for the potential given by (\ref{pts}), it is obtained
\begin{equation}\label{cnc}
    \partial_{\mu}J^{\mu}=-2i\bar{\phi}\gamma^{0}A(x)\phi\, ,\qquad \mathrm{for}\quad \mu=0,1\,.
\end{equation}

\noindent The current is not conserved and it is proportional to $A(x)$. In the same way, it is also useful to analysis the
orthogonality statement and it becomes
\begin{equation}\label{co2}
    \left(E_{k}-E_{k^{\prime}} \right) \int{dx\bar{\phi}_{k^{\prime}}\gamma^{0}\phi_{k}} +i \left[\bar{\phi}_{k^{\prime}}\gamma^{1}\phi_{k}\right]^{\,\,\,\,\infty}_{-\infty}
    -2\int{dx\bar{\phi}_{k^{\prime}}\gamma^{0}A(x)\phi_{k}}=0
\end{equation}

\noindent If one imposes the same boundary conditions (\ref{orto5}) , one gets
\begin{equation}\label{co3}
    \left(E_{k}-E_{k^{\prime}} \right) \int{dx\bar{\phi}_{k^{\prime}}\gamma^{0}\phi_{k}}=
    2\int{dx\bar{\phi}_{k^{\prime}}\gamma^{0}A(x)\phi_{k}}
\end{equation}

\noindent in this case the orthogonality statement is inconsistent. On the other hand, if one imposes the boundary conditions
\begin{equation}\label{co4}
    i \left[\bar{\phi}_{k^{\prime}}\gamma^{1}\phi_{k}\right]^{\,\,\,\,\infty}_{-\infty}
    =2\int{dx\bar{\phi}_{k^{\prime}}\gamma^{0}A(x)\phi_{k}}
\end{equation}

\noindent one gets the orthogonality statement (\ref{orto6}) and then (\ref{orto8}) is obtained. Therefore, the orthonormal condition (\ref{orto8}) for potentials with $PT$ symmetry is guaranteed if and only if (\ref{co4}) is satisfied.

In summary, the problem of relativistic problem of fermions subject to a PT-symmetric potential in the presence of position-dependent mass was reinvestigated. We have investigated the effects of a PT-symmetric potential on the conservation of the four-current and the orthonormal condition for the eigenspinors. We showed that the presence of the PT-symmetric potential does not lead to a conserved current. We also showed that the orthogonality criterion is consistent depending on the suitable boundary condition. The eigenspinors are orthonormal if and only if (\ref{co4}) is satisfied. Finally, it is useful to shed some light on the interaction of neutral fermions for clarify misconceptions diffused in the literature. It is known that, the four-dimensional Dirac equation with a mixture of spherically symmetric scalar, vector
and anomalous magnetic-like (tensor) interactions can be reduced to the two-dimensional
Dirac equation with a mixture of scalar, vector and pseudoscalar couplings when the fermion
is limited to move in just one direction ($p_{y} = p_{z} = 0$) \cite{st}. In this restricted motion the scalar
and vector interactions preserve their Lorentz structures while the anomalous magnetic-like
interaction becomes a pseudoscalar. The anomalous magnetic-like coupling describes the interaction of neutral fermions with electric fields, hence the two-dimensional version of the anomalous magnetic-like interaction, i.e. the pseudoscalar interaction will be responsible for the interaction of neutral fermions in (1+1) dimensions \cite{an1}-\cite{lui}. In Refs. \cite{chun4} and \cite{chun} the pseudoscalar interaction is not used, but the authors stated that the problem is associated to neutral fermions. Therefore, based on this foundation the titles of Refs. \cite{chun4} and \cite{chun} do not agree with truth.

\begin{acknowledgments}
The author would like to thank Professor Dr. Antonio S. de Castro. Thanks also go to editor and referee for useful comments and suggestions. This work was supported by means of funds provided by CAPES.
\end{acknowledgments}

\end{document}